\documentstyle[preprint,psfig,aps]{revtex}
\begin{document}
\def\pipi{$\pi$-$\pi^*$}
\def\be{\begin{equation}}
\def\ee{\end{equation}}
\def\fix{{\bf fix}}
\title{Application of time-dependent density-functional theory \\
to electron-ion coupling in ethylene}
\author{G.F. Bertsch$^{(a)}$, J. Giansiracusa$^{(a)}$ and
K. Yabana$^{(b)}$}
\address{$^{(a)}$Department of Physics and Institute for Nuclear Theory,\\
Box 351560,
University of Washington,
Seattle, WA 98915\\
$^{(b)}$Institute of Physics, University of Tsukuba,\\
Tsukuba 305-8571, Japan}
\maketitle

\begin{abstract}
To examine the applicability of the time-dependent density-functional 
theory (TDDFT) for treating the electron-nucleus coupling in excited states, 
we calculate the strength distribution associated with
the \pipi~transition in ethylene.
The observed optical transition strength at 7-8.5 eV region shows a 
complex structure arising from coupling to C-C stretch motion, 
to torsional motion, and to Rydberg excitations. 
The mean energy of the observed peak is reproduced to about 0.2  eV 
accuracy by the TDDFT in the local density approximation (LDA).
The reflection approximation is used to calculate the peak broadening. 
Roughly half of the broadening can be attributed to the fluctuation 
in the C-C coordinate.  The asymmetry in the line shape is also 
qualitatively reproduced by the C-C coordinate fluctuation.
We find, in agreement with other theoretical studies, that the 
torsional motion is responsible for the progression of weak 
transition strength extending from the peak down to about 6 eV.  The
LDA reproduces the strength in this region to about  factor of 3.
We conclude that the TDDFT is rather promising for calculating the
electron nucleus coupling at short times.  
\end{abstract}

\section{Introduction}
The time-dependent density functional theory (TDDFT) offers
a compromise between economy and accuracy for the calculation
of electronic excitations in atoms, molecules and condensed
matter.  The electronic excitation energies are typically
reproduced by a few tenths of an eV for states with large oscillator
strengths, and their strengths 
are typically reproduced within 25\% accuracy \cite{ya99,ca98}.
We are now interested in seeing how much further 
the theory can be applied at this level of accuracy and, 
in particular, how well it works in describing the coupling 
of electronic excitations to the nuclear degrees of freedom. 

In a previous publication, we examined this question taking the 
benzene molecule for a case study\cite{be01}. 
There we found that the TDDFT worked rather well 
at describing the nominally forbidden transitions, 
reproducing oscillator strength over 3 orders of magnitude 
with errors at the 30\% level.  
Besides giving transition strength to the symmetry-forbidden 
transitions, the electron-nucleus coupling induces a vibrational 
splitting of each electronic line.  While the detailed description
requires the Franck-Condon factors of the individual vibrational
states, the overall widths can be calculated with much simpler theory.
For benzene, we found that the overall widths associated with
the vibrational splittings were 
reproduced within 25\% accuracy.

The ethylene is an even simpler molecule whose absorption spectrum
has been subject to much theoretical study by the methods of 
quantum chemistry.  The observed photoabsorption spectrum in ethylene 
shows a strong, structured
peak in the region of 7-8.5 eV excitation that may be largely attributed
to the \pipi~ transition.  The studies
of ref. \cite{pe82} and \cite{me96b} present detailed calculations of
the vibronic couplings, but they do not make a direct comparison to
experiment.   We also note the calculations by
Ben-Nun and Martinez \cite{BM00} in which the subsequent 
nuclear motion is treated quantum mechanically.
The goal of the present article is less ambitious: 
we aim to see how well the density functional theory (DFT)
works for electron-nucleus coupling at short times, before the
system responds dynamically.
We will focus on the \pipi~ transition, which is of general 
interest for the optical properties of conjugated carbon systems.

The basic properties of the transition, besides its mean energy,
are its width (about 1 eV FWHM) and its oscillator strength,
measured to be $f=0.42$ \cite{co95}. 
A feature of particular interest is the long progression 
of strength on the low-energy side of the peak, extending down to about 
6 eV. 
We shall also apply the TDDFT to this feature and directly compare
the theoretical and the experimental strength 
in the region of 6-7 eV.
  
\section{Calculational Method}

We use the DFT to calculate the ground potential energy surface 
(PES) as well as the excited PES.  
In the present work, we use a simple density functional 
based on the local density approximation (LDA) \cite{pe81}.

Denoting the nuclear coordinates by $Q$, the ground PES
$E_{gs}(Q)$ is obtained in the 
static theory by minimizing the energy functional of the orbital 
variables. The optimized $\phi_i$ satisfy the Kohn-Sham equations
\be
-{\nabla^2 \over 2 m}\phi_i +{\delta {\cal E}\over\delta n}\phi_i=
\epsilon_i\phi_i.
\ee 
Here $\cal E$ is the energy functional including Coulomb interactions
but not including the electrons' orbital kinetic energy. 

The excited states are calculated by solving the linearized equations
of the TDDFT.  These equations are very close to the RPA equations,
differing only by the presence of an exchange-correlation term in the 
interaction.  For each ground state orbital $\phi_i$, 
there are two excited wave function, $\phi^+_i and \phi^-_i$.  The 
equations they satisfy are
$$
-{\nabla^2 \over 2 m}\phi_i^{\pm} +{\delta {\cal E}\over\delta
n}\phi_i^{\pm}-\epsilon_i\phi_i^{\pm} +{\delta^2 {\cal E}\over \delta n^2}
\delta n \phi_i=
(\epsilon_i\pm\omega)\phi_i^\pm,
$$ 
with the normalization condition,
$\langle\phi_i^+|\phi_i^+\rangle-\langle\phi_i^-|\phi_i^-\rangle=1$. 
The physical quantities extracted from the solutions are the eigenvalues 
$\omega$, representing vertical excitation energies, and the transition 
densities $\delta n = \sum_i\phi_i(\phi_i^++\phi_i^-)$ which represent 
the matrix elements of the density operator between the ground and the 
excited states. The excited PES is given by the sum of the ground PES 
and the excitation energy $\omega$,
\be
E_{ex}(Q) = E_{gs} (Q) + \omega (Q).
\label{eex}
\ee

To describe the oscillator strength distribution, we treat nuclear motion
classically except for the zero-point motion on the ground PES.
This is called the reflection approximation \cite{he78,bo00}.
Also, we neglect couplings between different vibrational
degrees of freedom and treat the fluctuations in various coordinates
$Q$ independently. 
The reflection formula for the strength function is 
\be
\label{reflection}
{d f\over d E} = f(Q) \left({d E_{ex} \over d Q}\right)^{-1} |\Psi(Q)|^2,
\ee
where the coordinate $Q$ in this expression is obtained by inverting 
the equation $E=E_{ex}(Q)$.  In the formula, $\Psi(Q)$ denotes
the vibrational nuclear wave function in the ground PES, 
normalized as $\int |\Psi|^2 d Q = 1$.  Also $f(Q)$ denotes the
oscillator strength at a fixed $Q$ between the 
ground and the excited state.
In ref. \cite{be01}, we used an even simpler approximation, 
namely taking the vibrations to be harmaonic and treating other factors
in eq.~(\ref{reflection}) as constant.  
Then the width is Gaussian with an energy variance $\sigma_E$
given by 
\be
\label{simple}
\sigma_E = \left.{d E_{ex} \over d Q}\right|_0 \sigma_Q,
\ee
where  $\sigma_Q$ is the variance of $Q$ in the ground state vibrational 
wave function. For a Gaussian profile, the variance is related 
to the width $\Gamma$ by  $\Gamma = 2.3 \sigma_E$.

Our computational method is quite different from usual quantum 
chemistry methods in that we employ a three-dimensional Cartesian 
grid representation for the orbital wave functions.  
Only valence 
electrons are explicitly included in the functional. More tightly bound 
electrons are taken into account by pseudopotentials constructed with a commonly 
used prescription \cite{tr91,kl82}.  Further details on the computation 
may be found in our previous publications, e.g. \cite{ya99}.
The important numerical parameters in the computations are 
the mesh spacing $\Delta x$ and the shape and size of the volume
on which the grid is defined.  In our previous work on conjugated
carbon systems, we used a grid spacing of $\Delta x = 0.3$ \AA.
With that mesh, orbital energies are converged to better than 
0.1 eV.  One can use the same coarse mesh to compute the coupling
to the vibrational degrees of freedom, provided the ground PES
is independently known.  In ref.~\cite{be01}
we only considered small amplitudes of nuclear motion,
obtaining  the necessary information from empirical data on the vibrations.
Here we want to consider larger amplitudes of motion, beyond the harmonic
regime. This requires a more accurate calculation of the ground PES, 
obtainable with a finer mesh.

\section{Results}

\subsection{Ground PES}

We first consider the ground PES. 
To obtain converged results for equilibrium nuclear positions
and vibrational frequencies, we found it necessary to use mesh spacing of
$\Delta x =0.2$ \AA. The orbital wave functions are defined on the grid
points that lie within a sphere. We found that a sphere of radius 
$R=5$~\AA~is adequate to
obtain converged ground orbital energies. Some properties related to the ground PES 
are shown in Table I.  
The C-C equilibrium distance is reproduced to 1\% accuracy.  
The curvature of the potential energy surface for the two 
important modes, the C-C stretch and the torsion, control 
the respective vibrational frequencies.  As may be seen from
the Table, the  empirical vibrational frequencies \cite{se89} 
are reproduced to about 5\% accuracy.  We also quote
the results of the {\it ab initio} calculations of ref.
\cite{BM99} and \cite{me96b}, which show more or less the
same accuracy.  We conclude that
the quality of the DFT results are encouraging to proceed with
the calculation of anharmonic and large amplitude fluctuation effects.

\subsection{Vertical Excitations}

Next we turn to the TDDFT excitation energies at the equilibrium
geometry (call the Franck-Condon point, FCP).  The TDDFT predictions
with the LDA energy functional are shown in Table II.
The \pipi~transition
appears at 7.5 eV excitation energy with an oscillator strength 
of $f=0.30$.  This energy is about 0.2 eV lower that the center of
gravity of the observed peak.  Also, comparing to the measured strength of 0.42 in this 
region, we infer that additional electronic transitions are likely
to be present.  Indeed, the TDDFT produces other states
in the same region of excitation. 
The three additional states have a Rydberg character with 
diffuse orbitals. Only the lowest of these, which is $s$-like
and has no angular nodes at large distance, is optically active.
It has a transition strength of $f=0.08$.  Adding this 
to the \pipi~ strength, the total strength is only 0.04 units abelow the
observed value.  We note that our oscillator strength are quite close
to those obtained by quantum chemistry methods, eg. ref. \cite{me96b}.
In principle, the LDA cannot give reliable excitation energies for
the Rydberg states, due to an incorrect asymtotic potential.  
However, the present results are within 0.3 eV of the more refined
theoretical values in the last column.  It might be possible to 
improve the energies by using a GGA density functional as in
ref.~\cite{na01},
but for the present study we will use
the LDA and confine our attention to the \pipi~transition.

\subsection{Coupling to the nuclear coordinates}

The nuclear coordinates that have the largest effect on the 
\pipi~transition are the C-C stretch coordinate and the torsion angle coordinate.
We shall consider both of these in the sections below.  The CH$_2$
wagging coordinates also play a role, first in mixing the \pipi~ excitation
with Rydberg excitations \cite{ry95,pe82}, and then at large torsion
angles providing a direct path from the excited PES to the ground PES
\cite{BM00}.  Since we ignore the Rydberg states and do not treat the
the multidimensional character of the nuclear coordinate space at all,
we shall neglect the wagging and other modes that couple indirectly.®

\subsubsection{C-C stretch motion} 

\begin{figure}
  \begin{center}
    \leavevmode
    \parbox{0.9\textwidth}
           {\psfig{file=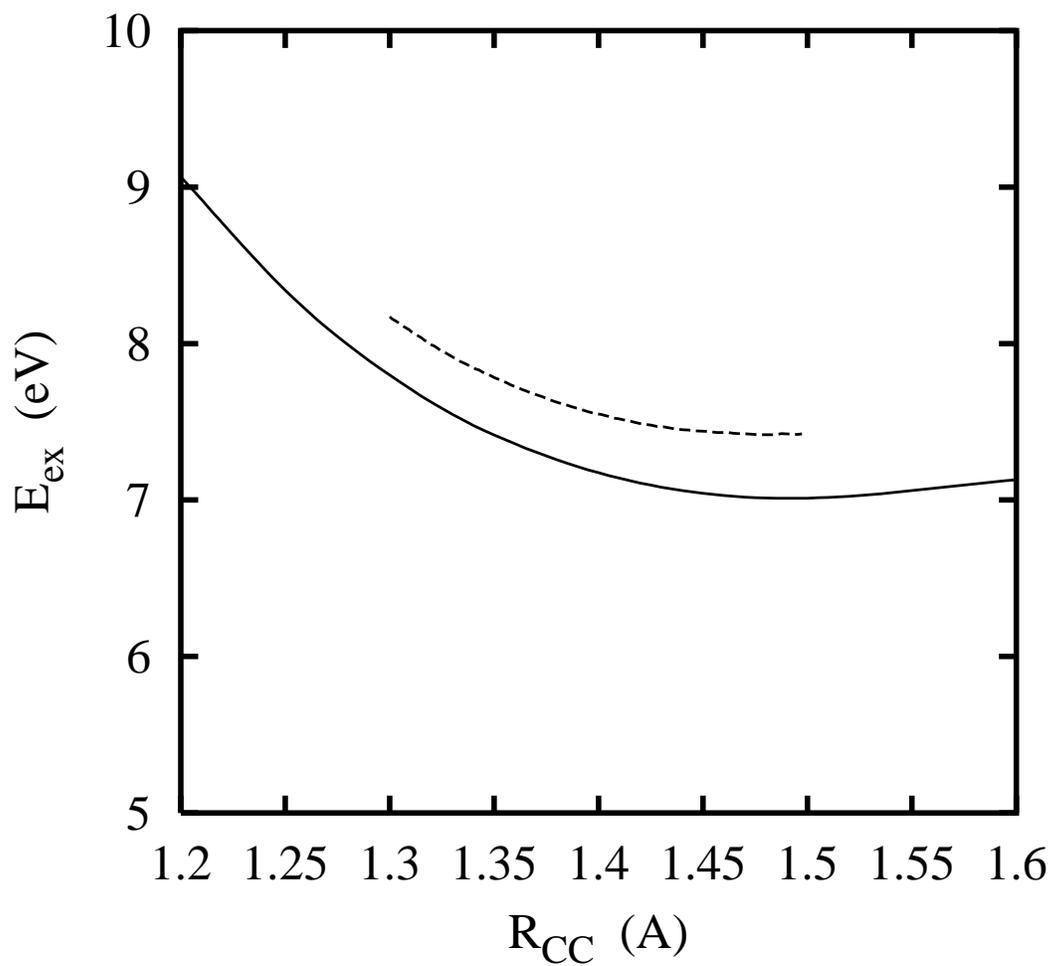,width=0.9\textwidth}}
  \end{center}  
\caption{Excited PES along the C-C stretch coordinate. Solid:  TDDFT;
dashed:  ref. [6].
}
\label{cc_eex}  
\end{figure}

Historically, the first candidate for the source of the width of the \pipi~
transition was the coupling to the C-C stretch coordinate.  Calculating
the excited PES along the stretch coordinate according to eq.~(\ref{eex}).
we find the results shown in Fig.~\ref{cc_eex}.  The excited 
PES has a minimum at
$R_{CC} = 1.5$ \AA~with an energy 7.0 eV.  The slope of the PES at the
FCP is given by $d E_{ex} / d Q = 8$ eV/\AA.  This is all we need to
estimate the width using eq.~(\ref{simple}).  Taking 
$\sigma_Q=0.04$~\AA~from empirical vibrational frequency, we find a 
variance of $\sigma_E\approx 0.3$ eV for the absorption peak.  
The corresponding width is $\Gamma = 0.7$ eV, roughly 2/3 of the 
observed peak width. We expect that the widths from fluctuation of 
different coordinates add in quadrature.  Thus, the width due to 
fluctuation in the C-C distance must be augmented by a similar 
width from other coordinates.

\begin{figure}
  \begin{center}
    \leavevmode
    \parbox{0.9\textwidth}
           {\psfig{file=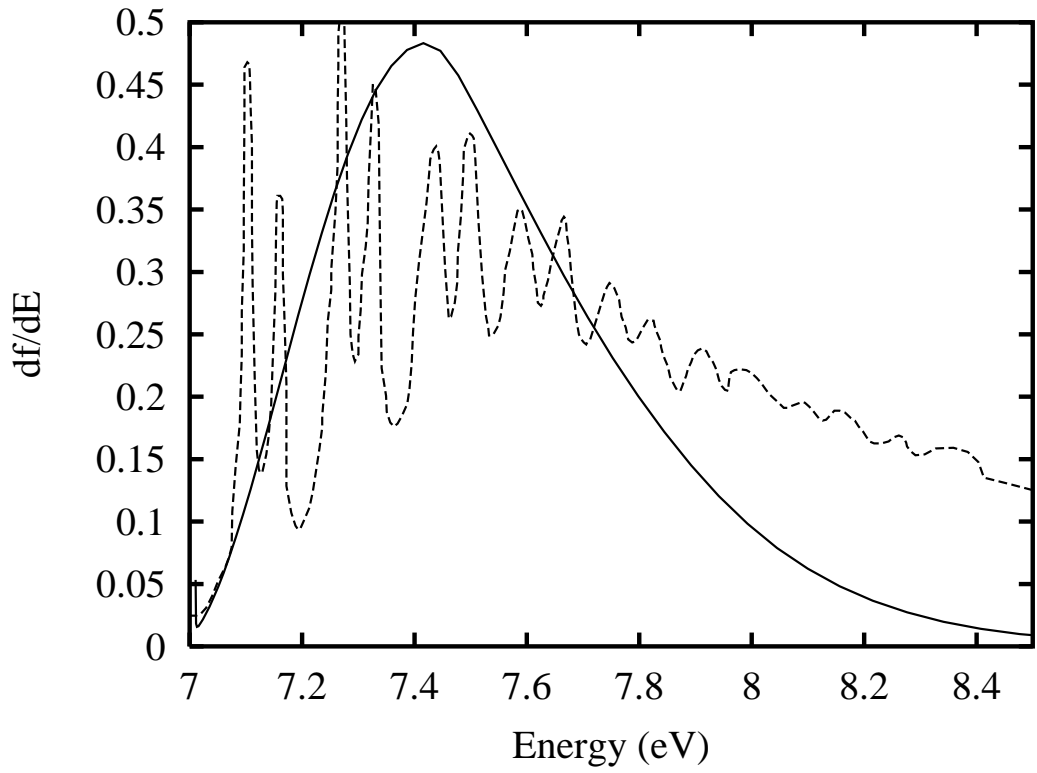,width=0.9\textwidth}}
  \end{center}  
\caption{Line broadening of the \pipi~excitation due to the zero-point motion
in the C-C bond coordinate.  Solid:  TDDFT theory; dashed: experimental
[13].
}
\label{cc_shape}  
\end{figure}

The observed peak has an apparent asymmetry, falling off more steeply 
on the low-energy side.  In principle the asymmetry can be calculated 
using the full expression eq.~(\ref{reflection}).  
We first solve the Schroedinger equation on the ground PES to 
get the wave function $\Psi(Q)$, and then apply eq. (3) directly.   
The result is shown 
in Fig.~\ref{cc_shape}. Indeed, a significant asymmetry is 
predicted, reflecting the anharmonicity in the ground PES and
the curvature of the excited PES.  
However the width is still too small.  We mention that the 
calculation of ref.~\cite{BM99} also predict an asymmetric peak.  
Their width includes all the vibrational modes, and comes out somewhat 
larger than the observed width.

\subsubsection{Coupling to torsional motion}

\begin{figure}
  \begin{center}
   \leavevmode
    \parbox{0.9\textwidth}
           {\psfig{file=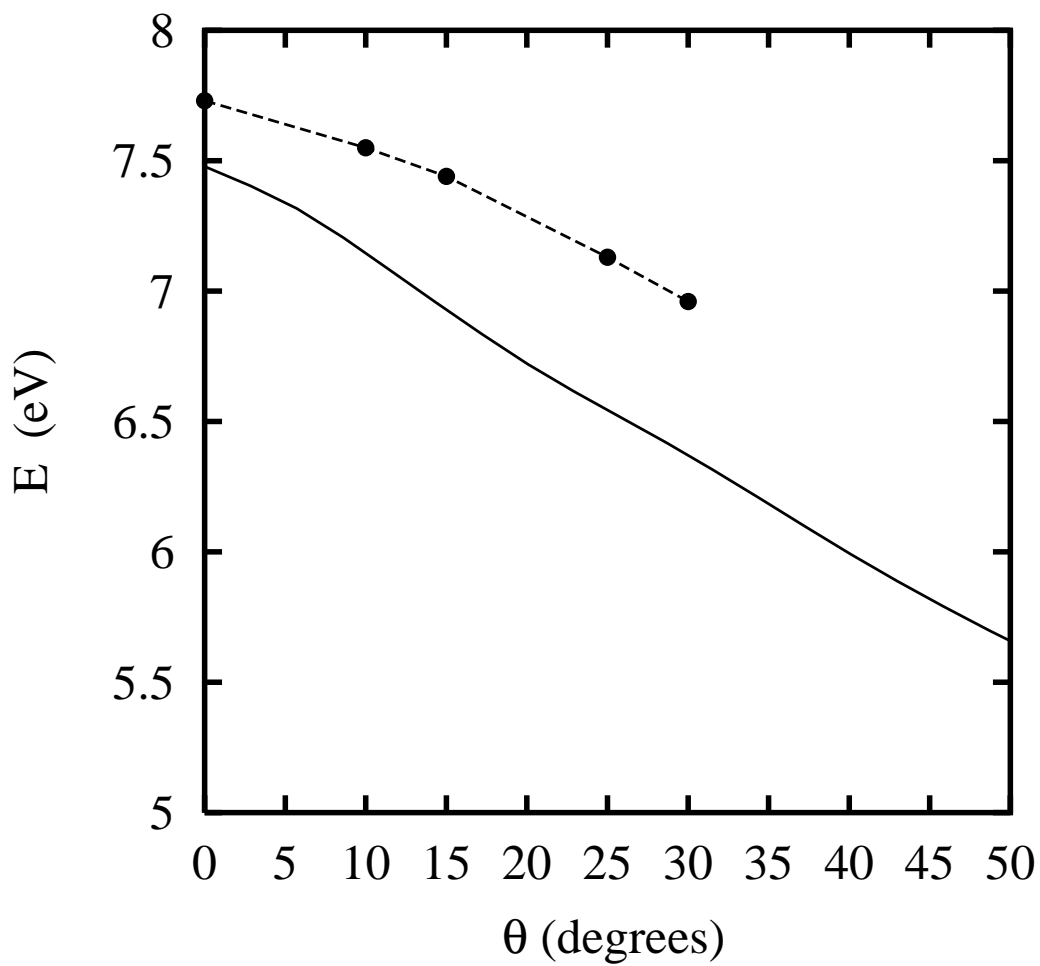,width=0.9\textwidth}}
  \end{center} 
\caption{Excited PES along the torsional coordinate. The solid line is the
TDDFT result with other coordinates
fixed at the FCP. The dashed line is from ref. [18].
}
\label{torsion_eex}  
\end{figure}

\begin{figure}
  \begin{center}
    \leavevmode
   \parbox{0.9\textwidth}
           {\psfig{file=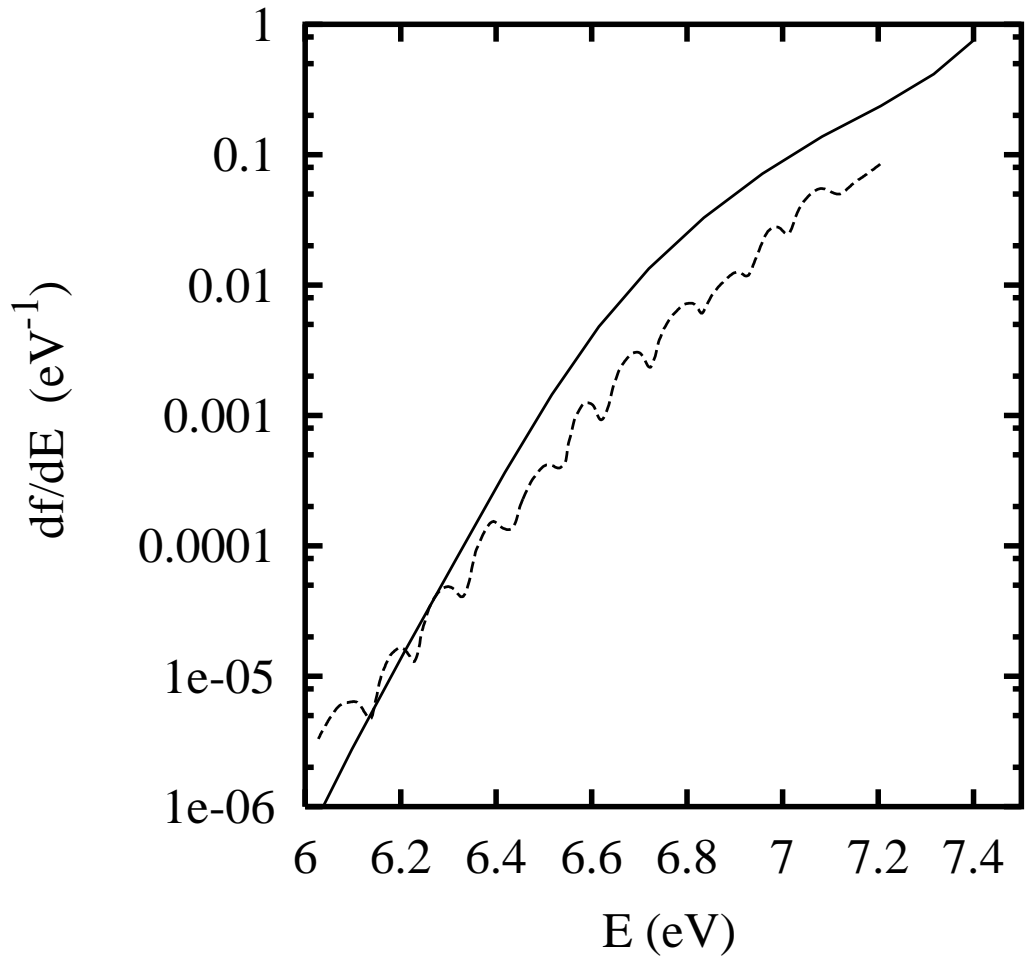,width=0.9\textwidth}}
  \end{center}  
\caption{Low energy absorption strength associated with the zero-point
motion in the torsional coordinate.  Solid: present theory; dashed:
experiment [19].
}
\label{torsion_shape}  
\end{figure}

Experimentally, the strong absorption peak at 7-8 eV has a progression
of strength at low energies, seen down to nearly 6 eV \cite{wi55}.
The C-C stretch PES doesn't drop to that low an energy,
leaving the 
torsional mode as the most likely source of the strength.
Combining the inertia $I_\theta$ (with $\hbar^2/I_\theta \approx 0.0048$
eV/r$^2$) of the torsional
motion with empirical vibrational frequency $\omega_{torsion}$, 
the zero-point fluctuation variance in the torsional angle is given by
$\langle\theta^2\rangle^{1/2} = \sqrt{\hbar/2 I_\theta \omega_{torsion}}\approx
8^\circ$.  
This is rather soft, and suggests that coupling to
this coordinate could be significant.  Unfortunately,
the \pipi~ excitation mixes with the Rydberg states at finite torsional
angles\cite{pe82,ry95}, splitting the original \pipi~ 
transition into several components.  Since the
Rydberg excitation energies are not accurate in the LDA,
we choose to ignore the splitting of the \pipi~state, taking
the diabatic state at the mean excitation energy associated
with an excitation along the $x$ coordinate.
The diabatic excited PES is shown in Fig.~\ref{torsion_eex}.  
Using it in eq.~(\ref{reflection}), we obtain the strength
distribution shown in Fig.~\ref{torsion_shape}. 
Qualitatively the measured fall-off is reproduced, but in detail 
the predictions are not as accurate as we had found for the 
forbidden transitions in benzene. 
In the region of  6.5-7.0 eV, the theoretical strength is too 
high, by about a factor of three.  The agreement is much better
at lower energies, down to 6.1 eV, the lowest energy measured
in ref.~\cite{wi55}.  However, the theory should be more reliable
the closer to the FCP, so the low energy agreement is probably a 
fortuitous cancellation of opposite-sign errors.

A likely cause of the discrepancy just below
7 eV is the error in the TDDFT excitation energy.
In Fig.~\ref{cc_shape}, we saw that the strength around 
the main peak in the LDA calculation is somewhat lower than the observed one. 
This is within a typical error of the TDDFT for excitation energy.
A overall shift of 0.2 eV in the excited state PES would improve 
the description of strength function both in the peak region and 
in the region as far down as 6.5 eV.
Shifting the excited state PES in that way, on the other hand, would 
destroy the present agreement below 6.5 eV. However, this large-angle region of
PES is only accessible by a deep tunneling of the nuclear motion.
It might be that the naive reflection treatment becomes inaccurate.
For example, it might be that the tunneling in the excited state 
vibrational wave function becomes important. It may also be that 
thermally promoted tunneling on the ground PES becomes significant 
in the deep tunneling region, even though the thermal excitation 
probabilities are small.  However, we leave it for future work to 
investigate these possibilities.

\section{Dynamics on the excited PES}

We now discuss briefly about nuclear dynamics on our excited PES.
An interesting measurement of a dynamic time scale on the excited
PES was reported in ref.~\cite{fa98}, made by a two-photon pump-probe 
technique.  In the experiment the two photon energies were 6.2 eV 
and 4.64 eV, and the average time delay between them could be varied.  
The ionization rate was measured as a function of the delay time.  
The observed ionization yield peaked when the pulse associated 
with the lower energy photon was delayed by about 50 fs from the 
pulse of the more energetic photon. Clearly, the ethylene must first 
be excited by the 6.2 eV photon before it can absorb the lower energy 
photon.  The experiment was interpreted in terms of lifetime of 
the state excited by the 6.2 eV photon, with an extracted value of
$30\pm15$ fs.  The ionization threshold of ethylene is 10.5 eV, 
so the total energy available is only 0.3 eV above the threshold.  

While the complete dynamics on the excited state PES are beyond
the TDDFT, the PES we have already constructed can be used to
estimate the accelerations in the short-time domain after excitation.  
According to the calculation in the previous section, 
the absorption strength at 6.2 eV comes from the torsional 
zero-point fluctuation
the hydrogens rotated to a angle of about $35{^\circ}$.  
Let us assume that the ionization potential of the molecule is 
independent of the angle.  Then, as long as the excitation energy is 
kept in the electronic degrees of freedom, the ionization can take 
place.  However, the gradient in the PES will cause an
acceleration in that coordinate, converting energy from potential
to kinetic.  The kinetic energy associated with nuclear motion
cannot be used in the ionization process, so these accelerations
give the first quenching of the ionization rate.  To make a
concrete estimate, we use the slope of the TDDFT PES to determine
the acceleration.  At $35{^\circ}$, the PES is still rapidly
falling, with a slope given by $d E_{ex}/ d\theta \approx 2$ eV/rad$^2$.
Taking this as the force in Newton's equation, the system
acquires a angular velocity given by
$$
{d \theta\over d t } = {1\over I_{\theta}} {d E_{ex} \over d \theta} t
$$
with a corresponding kinetic energy
$$
K = {1\over 2} I_{\theta} \left({d \theta \over d t}\right)^2.
$$
Evaluating these expression numerically, we find that it only takes
5 fs for the kinetic energy to grow to 0.5 eV.  This would put
the potential energy below the ionization threshold, and indicates
that the quenching should happen very rapidly.  

However, the fact that vibronic structure is seen in the strength
function tells one that the motion along the torsional coordinate 
is not very dissipative.  The system can accelerate to very large 
angle, bounce, and return to a state with no kinetic energy 
in the nuclear motion. Since we cannot say what the true 
irreversible rates are, the considerations here can only give a 
lower bound on the effects of a time delay.  We note that the
calculation of ref.~\cite{BM00} treats the large angle motion
as well and obtains results consistent with the observed time correlation.[A

\section{Conclusions}

In this article, we aimed to elucidate how well the TDDFT works 
in describing electron-nucleus coupling. For this purpose, we 
analyzed the \pipi~ transition of ethylene molecule in the 6-8 eV 
region. The DFT and TDDFT were employed to construct the ground and 
excited PES in the LDA. A simplified treatment in the reflection 
approximation is made for the Frank-Condon factor.
The TDDFT describes reasonably well the vertical excitation energy 
and the magnitude of oscillator strength.
The roles of the C-C stretch motion and the torsional motion on
the spectral line-shape were investigated. 
We found that the C-C stretch coordinate contributes substantially to the
width of the transition, but the torsional coordinate is responsible for the
low energy tail.  There the transition strength is reproduced by about a
factor of three over 4 orders of magnitude.  
We thus conclude that the TDDFT survives the first tests as a useful
approximation to treat electron-nucleus couplings in excited states.
As next steps, a more sophisticated treatment of
the quantum nuclear motion would be required, as well as an improved
treatment of the exchange-correlation potential beyond the LDA.

\section*{Acknowledgment}  
We thank T.J. Martinez for helpful discussions, and P.A.~Snyder for
correspondance leading to a corrected Table 2.  This work
was supported by the Department of Energy under Grant
DE-FG03-00-ER41132, and facilitated by the Computational
Materials Science Network.


\begin{table}
\caption{Ground state properties of ethylene.}
\begin{tabular} {ccccc}
& Experimental & DFT & Ref. \cite{BM00,BM99} & Ref. \cite{me96b} \\
\tableline
$R_{CC}$ (\AA)& 1.339 & 1.335 & 1.347  &\\
$\omega_{CC}$ (eV)& 0.201 & 0.19 & 0.21 & 0.22 \\
$\omega_{torsion}$ (eV) & 0.127 & 0.12 & 0.134 & 0.135 \\
\end{tabular}
\end{table} 

\begin{table}
\caption{Calculated electronic excitations in ethylene at the FCP.  The
lowest 4 excitations are mainly configurations made by exciting a
$\pi$ electron to the $\pi^*$ or to a Rydberg orbital. We follow the
convention used by ref. [13] for the $x,y,z$ axes.} 
\begin{tabular} {ccccc}
Excited orbital & $\omega$ (LDA) & f (LDA) & Ref. \cite{me96b}& Ref.
\cite{BMb00} \\
\tableline
$s$-Rydberg & 6.8 &0.08 & 7.3,0.08 & 6.8\\
$p_y$-Rydberg & 7.3 & 0 & &7.5 \\
$p_z$-Rydberg & 7.4 & 0 & &7.5 \\
$\pi^*$  & 7.5 eV & 0.30  &8.1,0.36& 7.8  \\
$p_x$-Rydberg & 7.8 & 0 & & \\
\end{tabular}
\end{table}


\end{document}